\documentclass[10pt,aps,pre,onecolumn,superscriptaddress,amsmath,amssymb]{revtex4-2}
\usepackage[utf8]{inputenc}
\usepackage{amsmath,amsthm}
\usepackage{graphicx}

\usepackage{bbold}
\usepackage{braket}
\usepackage{xcolor}
\newcommand{\bE}{\mathbf{e}}

\newcommand{\tr}[1]{\ensuremath{\operatorname{tr}\!{#1}}}

\newcommand{\bea}{\begin{eqnarray}}
\newcommand{\eea}{\end{eqnarray}}

\global\long\def\ga{\gamma} \global\long\def\de{\delta}

\global\long\def\ell#1{\theta_{#1}}

\global\long\def\la{\lambda} 
\global\long\def\si{\sigma}

\global\long\def\eps{\epsilon}

\global\long\def\ga{\gamma} \global\long\def\de{\delta}

\global\long\def\no{\nonumber}

\theoremstyle{remark}

\global\long\def\braket#1#2{\left\langle #1|#2\right\rangle }

\begin{document}

\title{Liouvillian Exceptional Points in Quantum Brickwork Circuits}
\author{Vladislav Popkov}
\affiliation{Faculty of Mathematics and Physics, University of Ljubljana, Jadranska 19, SI-1000 Ljubljana, Slovenia}
 \affiliation{Department of Physics,
  University of Wuppertal, Gaussstra\ss e 20, 42119 Wuppertal,
  Germany}
\author{Mario Salerno} \affiliation{Dipartimento di Fisica
  ``E.R. Caianiello'', and INFN Sezione di Napoli - Gruppo Collegato di Salerno,
  Universit\`a di Salerno, Via Giovanni Paolo II, 84084 Fisciano (SA),
  Italy}

\begin{abstract}
We demonstrate that Liouvillian exceptional points (LEPs), previously explored only in continuous Lindbladian dynamics, also emerge in discrete brickwork completely positive trace-preserving (CPTP) circuits.
By analytically solving a minimal two-qubit brickwork model, we identify the conditions under which discrete-time LEPs arise and show that they retain the hallmark square-root eigenvalue splitting and linear-in-time sensitivity enhancement.
These results establish a direct bridge between continuous non-Hermitian physics and discrete quantum-circuit architectures, opening a path toward the realization of exceptional-point-based sensing on near-term quantum processors.
\end{abstract}
\maketitle

\section{Introduction}
In the presence of dissipation, the unitary evolution of a quantum system is broken, the map becomes effectively contracting, and the system is driven toward a steady state.
Among the possible manifestations of such non-unitary evolution are \emph{exceptional points} (EPs)—parameter values where the Liouvillian superoperator becomes non-diagonalizable~\cite{Minganti-2019,Minganti-2020}.
Liouvillian exceptional points (LEPs) in continuous Lindbladian dynamics promise enhanced sensitivity due to their non-analytic dependence on perturbations, such as eigenvalue splitting $\propto \sqrt{\varepsilon}$ for a second-order LEP~\cite{Chen-2017-Nature,Hodaei-2017-Nature,Lai-2019-Nature}, analogous to the sensitivity enhancement observed in non-Hermitian photonics~\cite{Ozdemir-2019-NatMater,Huang-2022}.
LEPs have recently been demonstrated in continuous-time experiments~\cite{Murch-2022-PRL} and analyzed in few-qubit theoretical models~\cite{Popkov-2025}.
EP-like degeneracies can also appear in auxiliary non-unitary transfer matrices used to compute correlation functions in closed, dual-unitary circuits~\cite{Hu-2025}. In that last case, however, the EPs pertain to the mathematical construction rather than the physical evolution of the system.

Here we focus on the discrete-time setting most relevant for current quantum devices, where evolution proceeds stroboscopically through brickwork completely positive trace-preserving (CPTP) maps~\cite{Prosen-2018,Prosen-2019,Bertini-2019-PRL}.
While EPs in non-Hermitian maps and LEPs in continuous Lindbladians have been studied independently, the emergence of LEPs in brickwork CPTP circuits—and the fate of their associated sensing advantages—remain unexplored.
The discrete time-step and CPTP constraints could, in principle, alter the Jordan structure or suppress per-step signal amplitudes, challenging the persistence of LEP physics in circuit-based architectures.

In this Letter, we present the first analytic demonstration that two-qubit brickwork CPTP circuits can host Liouvillian exceptional points that retain their characteristic sensing behavior.
By constructing minimal steps reproducing a target Lindbladian to first order in $\Delta t$, we analytically determine the manifold of LEPs and quantify the eigenvalue-splitting scaling in the discrete-time regime.
Our results show that LEPs are a generic feature of stroboscopic CPTP dynamics, thereby bridging the gap between continuous Lindbladian evolution and the discrete operations implemented in all programmable quantum devices.

\begin{figure}[tbp]
   \centering
   \includegraphics[width=0.3\textwidth]{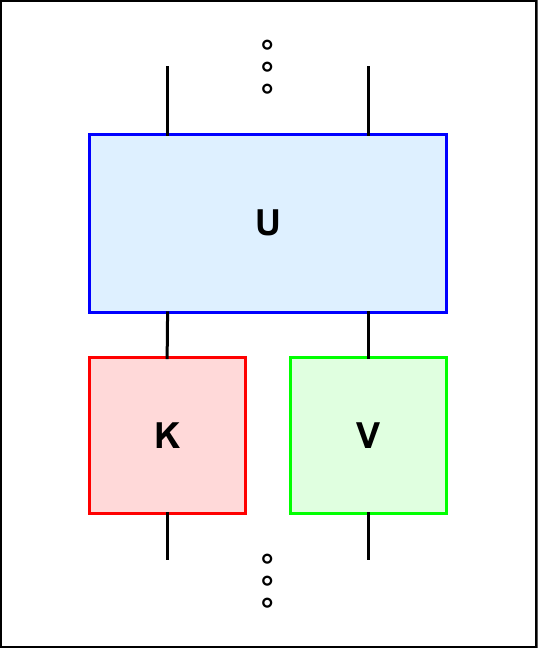}
   \caption{Schematic of the two-step brickwork protocol.  
   Even steps implement a two-qubit unitary gate $U$ (XXZ-type), while odd steps consist of a Kraus relaxation map on qubit 1 and a local unitary channel $V$ on qubit 2.  
   Periodic repetition of this protocol defines the brickwork CPTP map.}
   \label{FigSchema}
\end{figure}


\section{Model Equations} We consider a two-qubit brickwork circuit (Fig.~\ref{FigSchema}), composed 
of alternating unitary and dissipative layers.  Even steps apply a two-qubit gate
\begin{equation}
U=\left(
\begin{array}{cccc}
 1 & 0 & 0 & 0 \\
 0 & a & b & 0 \\
 0 & b & a & 0 \\
 0 & 0 & 0 & 1 \\
\end{array}
\right), \label{Ugate}
\end{equation}
parametrized by anisotropy $q$ and spectral parameter $\lambda$  as 
\begin{align}
&a\equiv \frac{q-q^{-1}}{q\lambda\!-\! (q\lambda)^{-1}},\quad
b\equiv \frac{\lambda-\lambda^{-1}}{q\lambda\!-\!(q\lambda)^{-1}}.\no
\end{align}
The gate acts on a pair of neighboring qubits via a folded map  as $\rho \rightarrow U 
\rho U^\dagger$. In the limit $\lambda \to 1+\delta$ with $\delta \ll 1$ it describes 
coherent time evolution of the density matrix $\rho$ via Trotter decomposition of $e^{-i H 
t}$ with $H$ the two-qubit XXZ Hamiltonian. Indeed, let us parametrize $q=e^{i \gamma},\, 
\lambda=e^\de$, with $\delta\ll 1$.  Expanding parameters $a$ and $b$ of $U$ to first 
order in $\de$ we get $a \approx 1 + i \de a_0 + O(\delta^2),\; b=-i \delta b_0 + 
O(\delta^2)$, with $a_0=\cot(\gamma)$, $b_0=1/\sin(\gamma)$, and U becomes:
\begin{equation}
U = 1 - i \delta \left(
\begin{array}{cccc}
 0 & 0 & 0 & 0 \\
 0 & -a_0 & b_0 & 0 \\
 0 & b_0 & -a_0 & 0 \\
 0 & 0 & 0 & 0 \\
\end{array}
\right) +O(\de^2) \approx \mathbb{I} - i\, \delta \, h_{12},
\end{equation}
where $h_{12}$ is a Hamiltonian density of the XXZ model,
\begin{equation}
h_{12}=J(\sigma_1^x \sigma_2^x+\sigma_1^y \sigma_2^y+ \Delta (\sigma_1^z \sigma_2^z - 
\mathbb{I}))\equiv h_{\rm XXZ}
\end{equation}
with $J=1/(2\sin\gamma),\, \Delta= \cos\gamma$. After $n$ iteration  steps,  with $n \de 
\equiv t$ fixed as $\de \rightarrow 0$, 
we have $U^n \rightarrow e^{-i t h_{\rm XXZ}}$,
so that the  density matrix undergoes a coherent unitary evolution, 
$\rho(t)=e^{-i h_{\rm XXZ} t}\rho(0)e^{i h_{\rm XXZ} t}$, under the XXZ Hamiltonian.

Odd steps act locally: qubit $1$ undergoes a dissipative Kraus map
\begin{align}
{\mathcal K}[\rho] = \sum_{j=1}^2 K_j \rho K_j^\dagger,\quad
K_1=\sqrt{1-\epsilon^2}\,\sigma^{+},\,
\;
K_2=\begin{pmatrix}1&0\\0&\epsilon\end{pmatrix},
\label{Kgate}
\end{align}
while qubit $2$ evolves under an arbitrary local unitary channel $\mathcal V[\rho]=V\rho V^\dagger$. The map (\ref{Kgate}) describes a  
relaxation  of a qubit towards the targeted state $\ket{\uparrow} \bra{\uparrow}$     
after a typical number of steps  $n_{\rm relax}=(-2/\log(\eps))$, 
where $0<\epsilon\le 1$ parametrizes the strength of the relaxation (see Appendix A).

For $0\leq \eps<1$ the Krauss map is strictly dissipative (non-unitary), while in the limit $\eps=0$
it reduces to a pure reset channel, realizable in current qubit-based quantum circuit experiments
\cite{GQAI_Science24a,GQAI_Nature22}. In this limit, any input qubit state $\rho$ is deterministically mapped to the fully polarized state
${\mathcal K}[\rho]=\ket{\uparrow}\bra{\uparrow}$, representing the most extreme form of non-unitary evolution (total dissipation).
In the opposite limit $\eps=1$, the Krauss map becomes unitary and coincides with the identity channel.
A full brickwork step is
\begin{equation}
\rho_{t+1}={\cal U} \rho_{t}= U\Big(\sum_{j=1}^2 (K_j\otimes V)\rho_t(K_j^\dagger\otimes V^\dagger)\Big)U^\dagger,
\label{eq:Evolution}
\end{equation}
or, in vectorized form, $\ket{\rho[t+1]}=\mathcal T\ket{\rho[t]}$, with discrete time Liouvillian
(or superoperator)
\begin{equation}
\mathcal{T} = (U \otimes U^*) \left( \sum_{j=1}^{2} (K_j \otimes V) \otimes (K_j^* \otimes V^*) \right).
\label{Superp}    
\end{equation}
In the $\Delta t\to0$ limit, this reduces  to a  continuous time Lindbladian  with jump 
operator $L=\sigma^+$ acting on the first qubit and XXZ Hamiltonian interactions 
(see Appendix A).

\section{Symmetry and block-diagonalization of the Liouvillian}

Both the Kraus gates and the two-qubit unitary $U$ preserve a $\mathbb{Z}_2$ parity 
symmetry,
$[\mathcal K,\Sigma_z]=[U,\Sigma_z]=0$, with $\Sigma_z=\sigma_z\otimes\sigma_z$.
For the full superoperator to inherit this symmetry we further require $[V,\sigma_z]=0$, 
in which case $V$ reduces to a pure $Z$ rotation.
The superoperator $\mathcal{T}$ can then be block-diagonalized by introducing the parity 
projectors
\begin{equation}
P_{\pm} = \tfrac{1}{2}\bigl(I_{16} \pm \Sigma_z \otimes \Sigma_z \bigr),
\end{equation}
which satisfy
$P_+ + P_- = I_{16}$,
$P_\pm P_\mp = 0$,
and $[P_\pm,\mathcal{T}]=0$.
Consequently,
\begin{equation}
\mathcal{T} = P_{+} \mathcal{T} P_{+} + P_{-} \mathcal{T} P_{-} \equiv \mathcal{T}_{+} + \mathcal{T}_{-},
\end{equation}
where $\mathcal{T}_\pm$ are rank-8 matrices acting on the even and odd parity subspaces, 
respectively.
After removing null rows and columns, the superoperator  takes the block-diagonal form
\begin{equation}
\mathcal{T} =
\begin{pmatrix}
\tau_+ & 0 \\
0 & \tau_-
\end{pmatrix}, \label{app:tauBlocks}
\end{equation}
where $\tau_\pm$ are $8\times8$ matrices  of the form 
\begin{equation}
\tau_\pm =
\begin{pmatrix}
A_\pm & B_\pm \\
C_\pm & D_\pm
\end{pmatrix},
\label{t-block}
\end{equation}
with  $A_{\pm}, B_{\pm}, C_{\pm}, D_{\pm}$,  $4\times4$ matrices given explicitly in 
Appendix~B,  most of which are sparse. For example, $A_{+}$ is diagonal, $B_{+}$ has a 
single nonzero element, and $C_{+}$ has only one nonzero row, while $D_{+}$ is a full 
matrix. Because of the sparsity of the off-diagonal blocks, the eigenvalues of $\tau_\pm$ 
can be efficiently computed using the Schur-complement discussed in the next section. 
Further simplifications lead to remarkably simple expressions for the superoperator 
spectrum (see below).

\section{Superoperator spectrum and the EP manifold} 
Although for $\mathbb{Z}_2$ symmetry the gate $V$ can in general be written as $V = e^{i 
\theta \sigma_z}$, to gain analytical insight into the spectrum and the location of 
exceptional points (EPs) we focus, in the following, on the case $V=I$ (i.e. $\theta=0$), 
which we refer to as the \textit{superintegrable} case for reasons given  below. 
This choice preserves the full dependence on $(\epsilon, q, \lambda)$ while considerably 
simplifying the secular equations of $\tau_\pm$, since in this limit the characteristic 
polynomials factorize into quadratic polynomials, as shown below. Within this setting, the 
entire spectrum can be obtained analytically, enabling a complete identification and 
classification of the non-Hermitian degeneracies that define the EP manifold of the 
brickwork superoperator.

In particular in  the easy-plane regime ($|q|=1$, $\lambda
\in\mathbb{R}$, $0<\epsilon\leq1$), the superoperator spectrum acquires a following simple 
structure: all eigenvalues are either analytic or involve a single square root. 
Explicitly,  the eigenvalues of $\tau_\pm$ blocks can be analytically computed using the 
Schur-complement formula
\begin{equation}
\det(\mu I_8 - \tau_\pm) = \det(\mu I_4 - A_\pm)\, \det\big(\mu I_4 - D_\pm - C_\pm (\mu I_4 - A_\pm)^{-1} B_\pm\big) = 0.
\label{schur}
\end{equation}

For the eigenvalues of $\tau_+$ the first factor of Eq.~\eqref{schur}, four eigenvalues of $\tau_+$ are directly
\begin{equation}
\mu_{1,2,3,4} = \{ 1, \epsilon^2, \epsilon, \epsilon \}.
\end{equation}
The remaining four eigenvalues follow from the second determinant in the RHS of (\ref{schur}) which factorizes as 
\begin{equation}
P(\mu) = (\mu - \epsilon )^2\, (\mu^2 + \xi \,\mu + \epsilon^2).
\label{polip}
\end{equation}
with $\xi$ given by 
\begin{equation}
\xi=\frac{\lambda ^2 q^4 \left(\epsilon
   ^2+1\right)+2 q^2 \left(\lambda ^4 \epsilon
   -\lambda ^2 (\epsilon +1)^2+\epsilon
   \right)+\lambda ^2 \left(\epsilon
   ^2+1\right)}{\left(\lambda ^2-q^2\right)
   \left(\lambda ^2 q^2-1\right)}. 
   \label{coeffp}
\end{equation}
Introducing the functions $f, Q$ as given in Eqs.~(\ref{eq:Q}),~(\ref{eq:Qf})
we have that the remaining eigenvalues can be written in the form: $\mu_{5,6} = \epsilon, \quad
\mu_{7,8} = \frac{(Q \mp f \lambda)^2}{4 (q^2 - \lambda^2)(\lambda^2 q^2 - 1)}$.
Similarly, the characteristic polynomial of $\tau_-$, after clearing common denominators, 
factorizes into four quadratic monic polynomials:
\begin{align}
P_1(\mu) &= \mu^2 + \; \frac{f \lambda}{\lambda^2-q^2}\,\,\mu + \;\epsilon \;\;\frac{\lambda^2 q^2 - 1}{q^2 - \lambda^2}, \\
P_2(\mu) &= \mu^2 + \epsilon\; \frac{f \lambda}{\lambda^2 - q^2}\,\mu + \epsilon^3\;\frac{\lambda^2 q^2 - 1}{q^2 - \lambda^2}, \\
P_3(\mu) &= \mu^2 +\; \frac{f \lambda}{1-\lambda^2 q^2}\,\mu + \epsilon \; \frac{ q^2-\lambda^2}{1-\lambda^2 q^2}, \\
P_4(\mu) &= \mu^2 + \epsilon \frac{f \lambda}{1-\lambda^2 q^2}\,\mu + \epsilon^3 \;\frac{q^2-\lambda^2}{1-\lambda^2 q^2},
\end{align}
with $f$ given in Eq.~(\ref{eq:Qf}). It is worth to note that the linear and constant coefficients of the polynomial $P_2$ are obtained from those of $P_1$ by multiplication with $\epsilon$ and $\epsilon^2$, respectively, and an analogous relation holds between the pairs $P_3, P_4$. This rescaling of monic quadratic polynomials implies that the roots of $P_2$ and $P_4$ coincide with those of $P_1$ and $P_3$, respectively, up to a factor $\epsilon$. Denoting $\mu_{9,\dots,16}$ as the eigenvalues of $\tau_-$, this implies:
$\mu_{11,12}=\epsilon\,\mu_{9,10}$,\quad $\mu_{15,16}=\epsilon\, \mu_{13,14}$, 
with the other eigenvalues following from $P_1$ and $P_3$ roots as:
\begin{equation}
\mu_{9,10} = \frac{f \lambda \pm Q}{2(\lambda^2 q^2 - 1)} \;\qquad \mu_{13,14} = \frac{f \lambda \pm Q}{2(q^2 - \lambda^2)}.
\end{equation}

\begin{figure}[tbp]
   \includegraphics[width=0.4\textwidth]{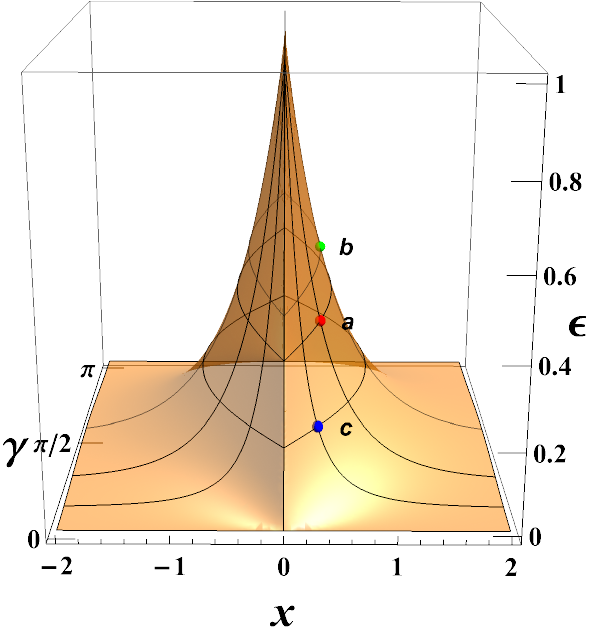}
  \caption{
Superoperator exceptional points manifold in parameter space 
$\{x\equiv \log \lambda, \gamma, \epsilon\}$ with $\epsilon\in[0,1]$ and $q=e^{i\gamma}$.  
Coordinates of representative points on the LEP manifold are:  
$a=\{0.3293,\pi/4,0.4\}$, $b=\{0.3466,\pi/2,0.5\}$, and $c=\{0.3013,\pi/9,0.2\}$.  
Bifurcation diagrams of $\mathcal{T}$ eigenvalues for EP $a$ are shown in Fig.~\ref{FigU1pi4}, while for those of EPs $b$ and $c$ are depicted in Fig.~\ref{Fig6}.
}
  \label{FigEPmanifold}
\end{figure}

In summary, the analytical expressions of the superoperator eigenvalues are:
\begin{align}
&\mu_1 = 1,\qquad \mu_{2} = \epsilon^2, \qquad \,\mu_{3}\,=\,\mu_{4}\,=\,\mu_{5} =\,\mu_{6} = \epsilon, \nonumber \\
&\mu_{7,8} = \frac{(Q \mp f \lambda )^2}{4 \left(q^2-\lambda ^2\right) \left(\lambda ^2
   q^2-1\right)},\quad
\mu_{9,10} = \frac{f \lambda \mp Q}{2(\lambda ^2 q^2-1)},  \label{eq:SuperoperatorSpectrum}\\
&\mu_{11,12}=\epsilon \mu_{9,10},\;\,
\mu_{13,14}=\frac{f \lambda \mp Q}{2(q^2-\lambda^2)},\;\,
\mu_{15,16} =\epsilon \mu_{13,14},\nonumber 
\end{align}
with
\begin{align}
Q &=\sqrt{\lambda ^2 (\epsilon -1)^2 (q ^4 + 1) + 2 q ^2 
\left(2 \lambda^4 \epsilon
    -\lambda ^2 (\epsilon +1)^2 + 2 \epsilon
   \right) },
   \label{eq:Q} \\
   f &= (q^2-1)(\epsilon+1). 
 \label{eq:Qf} 
\end{align}
\begin{figure}[tbp]
\centerline{\includegraphics[width=0.4\textwidth]{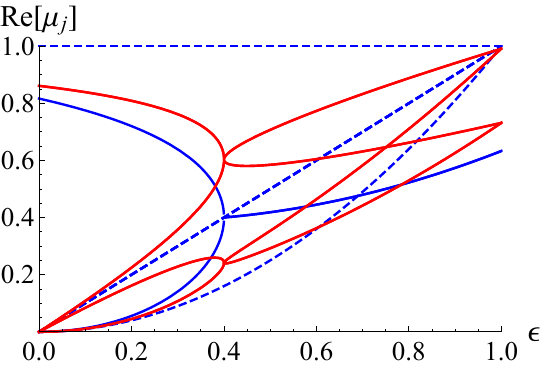}}  
\centerline{\includegraphics[width=0.4\textwidth]{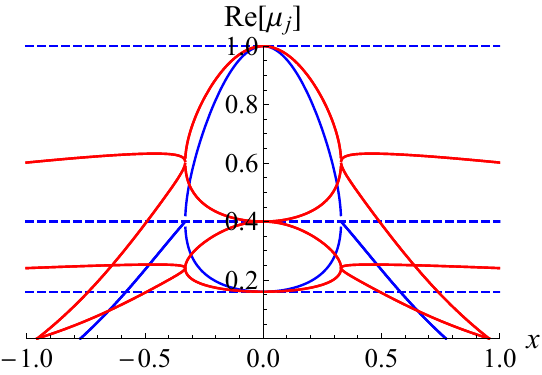}}  
  \caption{Real part of $\mathcal{T}$ eigenvalues as a function of $\epsilon$ for fixed $\lambda=1.39$ (top panel), and as a function of $\log \lambda$ for fixed $\epsilon=0.4$ (bottom panel).  
Parameters: $q=e^{i\pi/4}$, $V=I$. Red and blue curves correspond to $\mathcal{T}_+$ and $\mathcal{T}_-$, respectively,  while dashed lines refer to analytic (non-containing EP) eigenvalues $\la_j=1,\eps,\eps^2$.  
Bifurcation points occur at the EP (red dot), located on the EP manifold shown in Fig.~\ref{FigEPmanifold}.
 }
  \label{FigU1pi4}
\end{figure}

Denoting $q=e^{i\gamma}$, $\lambda=e^x$, one finds $Q=\sqrt{2}\lambda q \sqrt{F}$ with
\begin{align}
F = 2 \left((\epsilon -1)^2 \cos (2 \gamma )+4 \epsilon  \cosh (2 x)-(\epsilon +1)^2\right).
\end{align}
The zeros of $F$ correspond to second-order LEPs, i.e., coalescence of two eigenvalues into a Jordan block of size $2\times2$. Solving $F=0$ yields the EP manifold:
\begin{align}
\epsilon_{\rm EP} = 
\frac{ \cosh (2x)-\cos^2\gamma - \sqrt{2}|\sinh x|\sqrt{\cosh (2x)-\cos (2\gamma)}}{\sin^2
\gamma}.
\label{eq:EPpositionExact}
\end{align}
The function $\epsilon_{\rm EP}(x)$ is even in $x$ and has a cusp at $x=0$.  
EP  in the $3$-dimensional $\eps,\la,\ga$ space form a $2D$  surface, represented in 
Fig.~\ref{FigEPmanifold}. 
Bifurcation diagrams of the superoperator eigenvalues as functions of $\epsilon$ and of 
$x=\log(\lambda)$ for parameter values corresponding to the point $a$ and points $b,c$ of 
Fig~\ref{FigEPmanifold} , are depicted  in Figs.~\ref{FigU1pi4} and in Fig.~\ref{Fig6} of 
the Appendix B, respectively.

\section{Sensing properties on the EP manifold}

Liouvillian exceptional points are known to be associated with enhanced sensitivity to 
perturbations, originating from the non-analytic dependence of the Liouvillian spectrum on
external parameters. In particular, near a second-order LEP, small perturbations typically 
induce eigenvalue splittings scaling as $\sqrt{\varepsilon}$, leading to an amplified 
dynamical response compared to regular points in parameter space.

At an LEP, the discrete-time superoperator becomes non-diagonalizable. As a consequence of 
its Jordan-block structure, the stroboscopic evolution is no longer purely exponential but 
acquires polynomial prefactors, resulting in a contribution that grows linearly with the 
discrete time step $n$. 

This anomalous time dependence is the discrete-time counterpart of the well-known non-
Hermitian dynamical response at exceptional points and constitutes the fundamental 
mechanism underlying sensitivity enhancement in both continuous-time open quantum systems 
and non-Hermitian photonic platforms \cite{Minganti-2019, Chen-2017-Nature, Hodaei-2017-Nature, Ozdemir-2019-NatMater}.

To illustrate the impact of EPs on dynamics we consider the discrete-time evolution of 
observables.  For an initial state $\rho[0]$ the vectorized dynamics is
\begin{align}
\ket{\rho[n]} = \mathcal{T}^n \ket{\rho[0]},
\end{align}
with $\mathcal{T}$ the superoperator from Eq.~(\ref{Superp}).  
Away from EPs, $\mathcal{T}$ is diagonalizable and admits a biorthogonal set of left $\{\bra{w_j}\}$ and right $\{\ket{v_j}\}$ eigenvectors, $\braket{v_j}{w_k}=\delta_{jk}$.  
The expectation value of an observable $\hat g$ then reads
\begin{align}
\langle \hat g[n] \rangle = \sum_{j=1}^{16} (\mu_j)^n \, \gamma_j,
\quad 
\gamma_j = \alpha_j \, \mathrm{Tr}(\hat g\, \mathbf{v}_j),
\label{eq:ObservableEvolution}
\end{align}
where $\mu_j$ and $\mathbf{v}_j$ are the eigenvalues and eigenmatrices of $\mathcal{U}$, and $\alpha_j=\braket{w_j}{\rho[0]}$ are expansion coefficients.  
Enhanced sensitivity arises when $\hat g$ has overlap with eigenvectors associated with an EP.

The cleanest effect of EP can be best seen on observables which have significant overlap with the 
EP-containing eigenvectors of the superoperator ${\cal T}$, and by choosing an appropriate initial state. 
\begin{figure}[tbp]
   \includegraphics[width=0.41\textwidth]{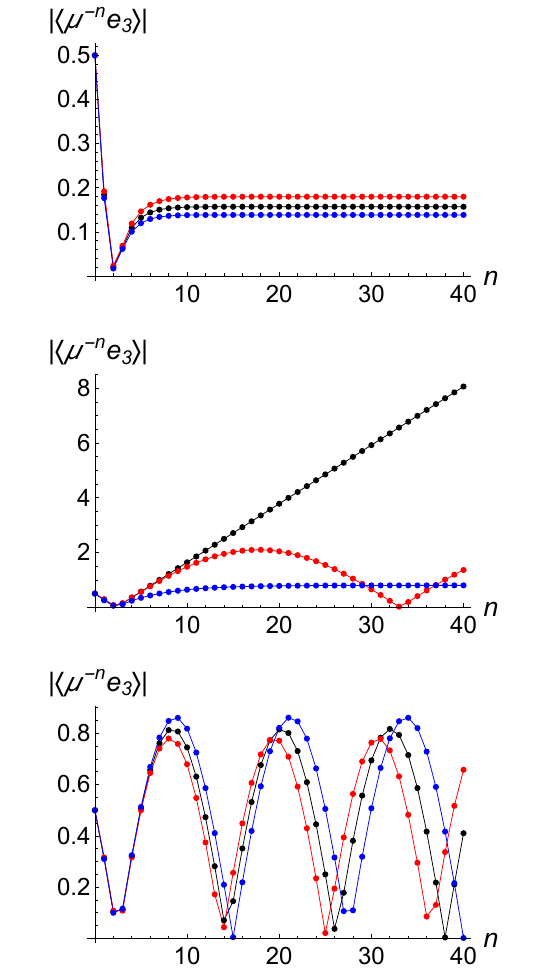}
  \caption{Renormalized value of $|\langle \hat \bE_3[n] \rangle|$
versus discrete time $n$,  below EP $\eps\equiv \eps_0 = 0.32, 0.32 \pm \de $ (upper row),  at EP   $\eps=0.4,  0.4\pm \de$ ( middle row),  and above EP   $\eps=0.48,  0.48\pm \de$ with $\de=0.01$.
 Black,  red and blue curves correspond to $\eps=\eps_0$,  $\eps_0 +\de $, $\eps_0 -\de $ respectively. 
Value of $\mu$ is chosen as $\mu = \max(|\mu_9(\eps)|,|\mu_{10}(\eps)|)$ at the respective $\eps$.
Linear increase of $|\mu^{-n}\langle \hat \bE_3[n] \rangle|$ at EP (black curve in the middle row)
is due to non-diagonalizability  of Liouvillian at EP,  see text.  Other parameters:
$q=\exp(i \pi/4),  \la = 1.39016$.  The EP is located at $\eps_{EP} = 0.4$ (point $a$ in Fig.~\ref{FigEPmanifold}).
 }
  \label{FigEP}
\end{figure}

As a concrete example we consider
 an initially pure quantum state $\rho[0] = \ket{\psi} \bra{\psi}$ with $ \bra{\psi} = 2^{-\frac12} (1,0,1,0)$ and 
take the  two-point  operator
\begin{align}
& \hat\bE_3 = \frac12 \si_1^{-} (I +\si_2^z),
\end{align}
whose dynamics involves only $\mu_9,\mu_{10}$:
\begin{align}
&\langle \hat \bE_3[n] \rangle = 
 \sum_{j=9,10}  (\mu_j)^n \ga_j  \label{eq:e3evolution}
\end{align}
with coefficients
\begin{align}
\gamma_{9}=\tfrac{g_{-}}{2(4+g_{-})}, \quad 
\gamma_{10}=\tfrac{g_{+}}{2(4+g_{+})}, 
\end{align}
and
\begin{align}
g_{\pm}=\frac{\big(\lambda(q^2-1)(\epsilon-1)\pm Q\big)^2}
{(\lambda^2-1)^2 q^2 \epsilon} 
\label{eq:gpm}
\end{align}
(for the explicit  derivation of the above equations see Appendix C).
The time  evolution of the quantity $|\langle \hat \bE_3[n] \rangle |$ properly rescaled,  is shown  on Fig.~\ref{FigEP},
at EP and its proximity. Everywhere except at the EP (black curve in the middle row),  the numerical data 
perfectly match the analytic expression  (\ref {eq:e3evolution}).
An additional factor $\mu^{-n} $ in front of the $\langle \hat \bE_3[n] \rangle $
is chosen so that to ensure that the quantity $|\mu^{-n} ]\langle \hat \bE_3[n] \rangle |$ enters a stable cycle for large  $n$. 
\begin{figure}[tbp]
   \includegraphics[width=0.4\textwidth]{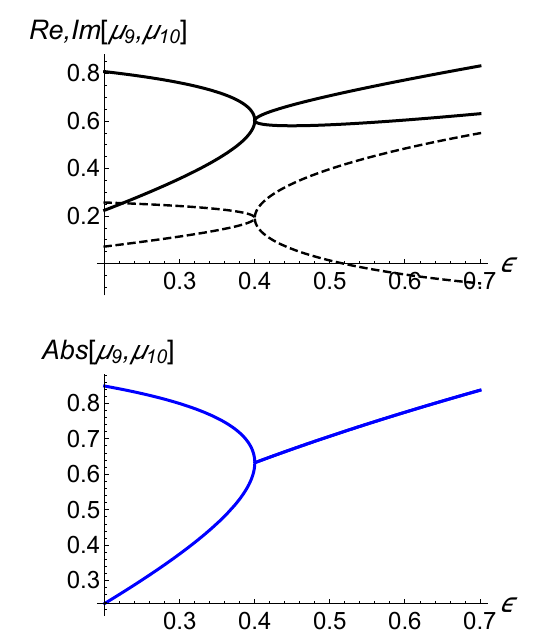}
  \caption{$\mu_9,\mu_{10}$,  entering Eq.~(\ref{eq:e3evolution})  versus $\eps$. 
  EP is located at $\eps_{EP} = 0.4$.  Bottom Panel shows absolute values $|\mu_9|,|\mu_{10}|$ while the Top 
panel shows their real part (solid lines ) and imaginary part (dashed lines).
Other parameters are fixed as in Fig.~\ref{FigEP}:
$q=\exp(i \pi/4),  \la \approx  1.39$.  
 }
  \label{FigEPeigen}
\end{figure}

Three different regimes can be seen clearly.  First, let us discuss the origin of the linear increase  at EP
(black curve,  middle row at Fig.~\ref{FigEP}).  
At the EP two eigenvalues coincide $\mu_9=\mu_{10}\equiv \mu_0$ and the respective $2 \times 2$ block $B$ of the 
Liouvillian acquires  Jordan form 
\begin{align}
& B=\left(
\begin{array}{cc}
 \mu_0 &1 \\
0& \mu_0
\end{array}
\right). \label{eq:Jordan}
\end{align} 
A repeated application of the above on a vector $\psi \in C_2$ gives  
\begin{align}
&B^n \psi =\left(
\begin{array}{cc}
 \mu_0^n &n \mu_0^{n-1} \\
0& \mu_0^n
\end{array}
\right)  \psi \label{eq:Jordan1}
\end{align} 
Assuming $\psi^T =(a,b)$ we obtain 
\begin{align}
&\mu_0^{-n} ||B^n \psi|| =\mu_0^{-n}  \left| \left| \left(
\begin{array}{cc}
 \mu_0 &1 \\
0& \mu_0
\end{array} \right)^n  \binom{a}{b}
\right| \right| \no \\
&=   n |\mu_0^{-1} b | \left( 1 + O\left( \frac{a}{ b n } \right)  \right), 
\end{align} 
e.g.  an asymptotic  linear increase with $n$,  seen on black curve,  in the middle Panel of Fig.~\ref{FigEP}.  
Note that 
if at the point $\eps_{EP} = 0.4$ the Liouvillian would have  conventional degeneracy,  i.e.  $B = diag(\mu_0, \mu_0)$,
then it would result in constant  in time $ \mu_0^{-n} ||B^n \psi|| = ||\psi|| = \sqrt{|a|^2+ |b|^2}$.   

At all the other points except EP we have diagonalizable Liouvillian,  with temporal evolution of the observable 
described by (\ref {eq:e3evolution}).  Notably, however,  there is an asymptotic constant
behaviour of  absolute value of $ \mu^{-n} \langle \hat \bE_3[n] \rangle $
below EP $\eps<\eps_{EP}$  and   asymptotic periodic behaviour above EP $\eps>\eps_{EP}$.  
To explain this,  let us take a look at the nature of the EP for $\mu_9,\mu_{10}$ in 
Fig.~\ref{FigEPeigen}:
Above EP,  absolute values of  $\mu_9,\mu_{10}$ are the same,  while 
below EP,  they are different.  This straightforwardly explains the qualitative differences 
in  Fig.~\ref{FigEP} for $\eps <\eps_{EP}$ and $\eps >\eps_{EP}$. 

 \section{Conclusion}
In this work, we have extended exceptional-point physics, traditionally formulated within continuous Lindbladian dynamics, to discrete architectures built from physical completely positive trace-preserving maps. In particular, we analytically demonstrated the existence of a full manifold of Liouvillian exceptional points in a minimal two-qubit brickwork CPTP circuit and showed that, in the vicinity of this manifold, the dynamics of observables exhibits the hallmark linear-in-time sensitivity enhancement. This establishes that discrete-time Liouvillian exceptional points preserve the essential dynamical features underlying the enhanced sensitivity known from their continuous-time counterparts.

From a practical perspective, operating a quantum circuit close to such an exceptional point implies that small perturbations of experimentally controllable parameters, such as gate angles or dissipative rates entering the CPTP maps, can induce parametrically enhanced changes in the evolution of the density matrix.

We note, however, that while the non-analytic spectral response near exceptional points leads to an amplified parameter dependence, a debate has emerged regarding whether this translates into a genuine metrological advantage in practical settings. Much of this discussion has focused on exceptional points in non-Hermitian Hamiltonians, often realized in post-selected or classical wave systems, where the enhanced susceptibility can be accompanied by increased sensitivity to noise, potentially offsetting the gain \cite{Ding-2023,Zeng-Nori-2025}. In this context, recent works have shown that non-Hermitian sensing schemes can outperform Hermitian ones under specific conditions, such as in the presence of technical noise, although such advantages do not necessarily rely solely on the presence of an exceptional point \cite{Arkhipov-Nori-2026,Yang-2026}.

While our analysis focuses on establishing the existence and structure of Liouvillian EPs in stroboscopic quantum dynamics, it naturally motivates future investigations of concrete sensing protocols in open quantum circuits.

\vskip .3 cm
\begin{acknowledgments} V.P.  acknowledges support by ERC Advanced grant
 No.~101096208 -- QUEST,  Research Program P1-0402 and Grant N1-0368 of Slovenian Research and Innovation Agency (ARIS) and by Deutsche Forschungsgemeinschaft through DFG project KL645/20-2. V.P. wishes to thank the Department of Physics of the University of Salerno, where this project initiated, both for hospitality and partial financial support.
\end{acknowledgments}

\bibliography{Brick2PRA}


\appendix

\section{Kraus map and continuous-time limit of the qubit circuit dynamics}  
The dissipative channel acting on one of the qubits is defined as $\mathcal K[\cdot] = \sum_{j=1}^2 K_j [\cdot] K_j^\dagger$
with Kraus operators
\begin{align}
K_1(\epsilon) = \sqrt{1-\epsilon^2}\, \sigma_1^{+}, 
\qquad 
K_2(\epsilon) =
\begin{pmatrix}
1 & 0 \\
0 & \epsilon
\end{pmatrix}.
\label{eq:Kraus}
\end{align}
This channel is completely positive and trace preserving by construction, as
$\sum_j K_j^\dagger K_j = \mathbb{I}$. The eigenoperators of $\mathcal K$ are 
$\psi_0 = \ket{\uparrow}\bra{\uparrow},\; 
\psi_1 = \sigma^z, \;\
\psi_2 = \sigma^+, \;
\psi_3 = \sigma^-$,
with corresponding eigenvalues
$\lambda_0 = 1,  \lambda_1 = \epsilon^2, \qquad \lambda_2=\lambda_3=\epsilon$.
These coincide with those of the channel generated by the Lindblad dissipator
$\exp(\Gamma t \,\mathcal D_{\sigma^+})$,
\begin{align}
\mathcal D_{L}[\rho] = 2 L \rho L^\dagger - \{L^\dagger L,\rho\},
\label{eq:Dissipator}
\end{align}
whose spectrum is $\mu_0=1,\;\; \mu_1=e^{-2 \Gamma t},\;\;  \mu_2=\mu_3=e^{-\Gamma t}$.
The equivalence $\mathcal K^n \simeq \exp(\Gamma t \,\mathcal D_{\sigma^+})$ follows in the scaling limit
$\epsilon = e^{-\Gamma t / n}, \; n\to\infty$.
The complete update of the density matrix over one step reads
\begin{align}
\rho_{t+1} = U \Bigg( \sum_{j=1}^2 (K_j \otimes V)\, \rho_t \,(K_j^\dagger \otimes V^\dagger)\Bigg) U^\dagger = \,{\cal U} \,\rho_t,
\label{eq:EvolutionSM}
\end{align}
where $U$ is the two-qubit fSim gate defined in Eq.~(1) of the main text and $V$ is an arbitrary $2\times 2$ unitary matrix.  

 In the weak coupling limit $\la=1+\eps$ with $\eps = (q-q^{-1})t/(i n)$ and $n$ very large,  we obtain
$U \approx I -i \frac{t}{n}  h$,  where $h =\si^x \otimes \si^x +\si^x \otimes \si^x +\frac{q+q^{-1}}{2} (\si^z \otimes \si^z - I)$
is the density of the XXZ spin chain Hamiltonian.  Then,  
using the Trotter formula, 
one obtains $\lim_{n\rightarrow \infty} {\cal U}^n \rho = e^{{\cal L}t} \rho $ where ${\cal L}$ is the Liouvillian dynamics of the boundary-driven Lindblad master equation
\begin{align}
\frac{\partial \rho}{\partial t} ={\cal L} [\rho]=  - i[H,\rho] + \Gamma  \mathcal D_{L_{\si_1^{+}}}[\rho],
\label{eq:LMESM}
\end{align}
where $H= \sum_{n=1}^{N-1} h_{n,n+1}$ is the  XXZ spin chain Hamiltonian.
Thus, the brickwork circuit can be seen as  a generalization of a Trotter discretization of Lindblad dynamics.

\section{Explicit form of the Liouvillian blocks}

In a convenient basis obtained by a similarity transformation within each parity sector,
the $4\times4$ blocks $A_\pm$, $B_\pm$, $C_\pm$, $D_\pm$  in 
Eq.~\ref{t-block} are:
\begin{equation}
\hspace{-.2cm}
A_+=\left(
\begin{array}{cccc}
 1 & 0 & 0 & 0 \\
 0 & \epsilon ^2 & 0 & 0 \\
 0 & 0 & \epsilon   & 0 \\
 0 & 0 & 0 & \epsilon  \\
\end{array}
\right),
\,
B_+=\left(
\begin{array}{cccc}
 0 & 1-\epsilon ^2 & 0 & 0 \\
 0 & 0 & 0 & 0 \\
 0 & 0 & 0 & 0 \\
 0 & 0 & 0 & 0 \\
\end{array}
\right),
C_+=\left(
\begin{array}{cccc}
0 & 0 & a_1(1-\eps^2) & 0  \\
0 & 0 & a_3(1-\eps^2) & 0 \\
0 & 0 & -a_2(1-\eps^2) & 0 \\
0 & 0 & a_2(1-\eps^2) & 0 \\
\end{array}
\right),
D_+=\left(
\begin{array}{cccc}
 a_1 & a_3 \epsilon^2 & -a_2 \epsilon & a_2 \epsilon
   \\
 a_3 & a_1 \epsilon^2 & a_2 \epsilon & -a_2 \epsilon
  \\
 -a_2 & a_2 \epsilon^2 & a_1 \epsilon & a_3 \epsilon \\
 a_2  & -a_2 \epsilon^2 & a_3 \epsilon & a_1 \epsilon \\
\end{array}
\right),
\nonumber
\end{equation}

\begin{equation}
\hspace{-0.4cm}
A_-=\left(
\begin{array}{cccc}
 b_3 & 0 & b_4 \eps & 0 \\
 0 & b_3 \eps^2 & 0 & b_4 \eps \\
 b_4 & 0 & b_3 \eps & 0 \\
 0 & b_4 \eps^2 & 0 & b_3 \eps \\
\end{array}
\right),\;
B_-=\left(
\begin{array}{cccc}
 0 & b_3(1-\eps^2) & 0 & 0 \\
  0 & 0 & 0 & 0 
  \\
 0 & b_4(1-\eps^2) & 0 & 0 
 \\
0 & 0 & 0 & 0
 \\
\end{array}
\right),
C_-=\left(
\begin{array}{cccc}
 0 & b_1(1-\eps^2) & 0 & 0 \\
 0 & 0 & 0 & 0 \\
 0 & 0 & 0 & 0 \\
 0 & b_1(\eps^2-1) & 0 & 0 \\
\end{array}
\right),
\,
D_-=\left(
\begin{array}{cccc}
b_1 & 0 & 0 & -b_2\eps \\
 0 & b_1 \eps^2 & -b_2\eps & 0 \\
 0 & -b_2 \eps^2 & b_1 \eps & 0 \\
-b_2 & 0 & 0 & b_1\eps \\
\end{array}
\right),
\nonumber
\end{equation}
with  coefficients $a_i, b_i$ given by:
\begin{eqnarray}
&& a_1 = \frac{\lambda ^2 \left(q^2-1\right)^2}{\left(q^2-\lambda ^2\right) \left(\lambda ^2 q^2-1\right)},\quad
a_2=\frac{\lambda  \left(\lambda ^2-1\right) q \left(q^2-1\right)}{\left(q^2-\lambda ^2\right) \left(\lambda ^2 q^2-1\right)},\quad
a_3=-\frac{\left(\lambda
   ^2-1\right)^2 q^2}{\left(q^2-\lambda ^2\right) \left(\lambda ^2 q^2-1\right)}, \\
&& b_1=\frac{\lambda  \left(q^2-1\right)}{q^2-\lambda ^2},\qquad\;
b_2=\frac{\left(\lambda ^2-1\right) q}{q^2-\lambda ^2},\qquad\;
b_3=\frac{\lambda  \left(q^2-1\right)}{\lambda ^2
   q^2-1},\qquad \;
b_4=\frac{\left(\lambda ^2-1\right) q}{\lambda ^2 q^2-1} .
\end{eqnarray}

\begin{figure}[h]
\centering
\includegraphics[width=0.8\columnwidth,clip]{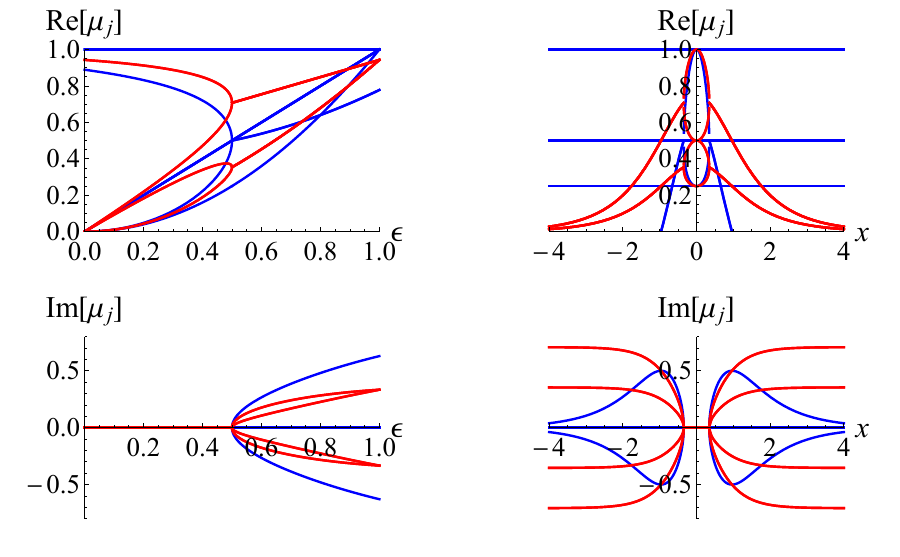}
\hspace{0.0\columnwidth}
\includegraphics[width=0.8\columnwidth,clip]{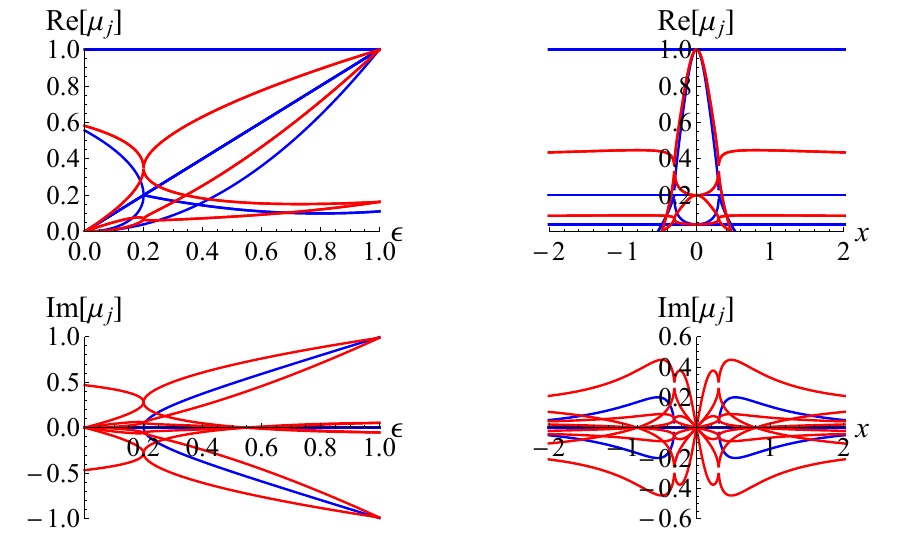}
\caption{
Real and imaginary parts of the superoperator eigenvalues corresponding to blocks
$\tau_+$ (blue color) and $\tau_-$ (red color) in (\ref{app:tauBlocks}),
versus $\epsilon$ (left column) and  $x=\log(\lambda)$ (right column),  for parameter values corresponding to points $b$ (first and second row panels)  and $c$ (third and fourth row panels) in Fig.~\ref{FigEPmanifold} in the main text. 
}
\label{Fig6}
\end{figure}

Note that most of the above  blocks are sparse, a property that enables an analytic evaluation of the eigenvalues of $\tau_\pm$ using the Schur-complement formula in Eq.~\ref{schur}. In Fig.~\ref{Fig6} we show various types of bifurcation diagrams  for the eigenvalues of the Liouvillian  
as obtained from the above block decomposition using Eq.~\ref{schur}.

\section{Time evolution of the observables: derivation of Eqs.~(16)-(18).}
\label{app:TimeEvolutionObservables}
The diagonalizability of the superoperator ${\cal T}$ in Eq.~4 implies  the existence of the 
biorthogonal set of its  left and right eigenvectors $\bra{w_j}$, $\ket{u_j}$,  corresponding to the eigenvalues $\mu_j$ of  ${\cal T}$ and satisfying $\braket{w_j}{u_k} = \de_{j,k}$.

The time evolution of an  observable $ \hat{g}$ is given by
\begin{align}
&\langle \hat g[n] \rangle = \tr(\hat g \, \rho[n])
\end{align}
where $\rho[n] = {\cal U}^n \rho[0]$  is a reduced density matrix of the system after $n$ steps. 
Using biorthogonal basis of the vectorized version ${\cal T}$ of the map ${\cal U}$ we write
\begin{align}
\ket{\rho[n]} 
=  \sum_{j=1}^{16}  (\mu_j)^n \braket{w_j}{\rho[0]}\ket{v_j}
\label{eq:ObsEvolution}
\end{align}
where $\ket{\rho[n]} $ is a vectorized density matrix after $n$ steps.  Using $\bra{g} $ for the vectorized version of $ \hat{g}$
we obtain 

\begin{align}
&\langle \hat g[n] \rangle =  \sum_{j=1}^{16}  (\mu_j)^n \braket{w_j}{\rho[0]} \braket{g}{v_j}
 \label{app:ObservableEevolution}
\end{align}
In order to see clearly the effect of EP,  we need  to choose $\rho[0]$ and $\hat g$ such that the major  contribution
in (\ref{app:ObservableEevolution}) comes from some of EP-containing eigenvalues  $\mu_7,\mu_8, \ldots , \mu_{16}$.

First, we list all the $\ket{v_j}$: using  standard notation $(e_{j})_k = \de_{jk}$, we have: 
for the regular analytic ${\cal T}$ eigenvalues $\mu_1,\mu_2, \ldots , \mu_{6}$ the eigenvectors are 

\begin{align}
&\ket{v_1} =e_1,   \label{eq:v1}\\
&\ket{v_{2}} = e_1-e_6-e_{11}+e_{16}, \label{eq:v2} \\
&\ket{v_{3,4}}=(1+\eps) e_1 +x_4 e_4 - \eps e_6 +F e_{10} -e_{11} + x_{13} e_{13}, \\
&\ket{v_{5,6}}=(1+\eps) e_1 +x_4 e_4 - \eps e_6 -F e_{7} -e_{11} + x_{13} e_{13}, \\
&F = \frac{(1-q^2)(\eps-1) \la} {q(-1+ \la^2)},
\end{align}
where $x_{4},x_{13}$ are arbitrary constants. 
The EP-containing eigenvalues  contain dependence on the square root  $Q$ from (\ref{eq:Q}), namely, 
\begin{align}
&\ket{v_7} =a_1 e_1 + \sum_{j=6,7,10,11} b_j(Q) e_j,   \label{eq:v7}
\end{align}
 $a_j$ are analytic coefficients while $b_j (Q)$ are linear in $Q$. 
Then,
\begin{align}
&\ket{v_8} = - \ket{v_7}+ 2Q (- \lambda  \left(q^2-1\right)) (e_{6} -e_{11})  + 2 Q \left(\lambda ^2-1\right) q (e_{7} -e_{10}).
\end{align}
$\ket{v_9},\ket{v_{10}}$ have especially simple form 
\begin{align}
&\ket{v_9} = e_5  -\frac{Q-\lambda  \left(q^2-1\right) (\epsilon -1)}{2 \left(\lambda ^2-1\right) q \epsilon } e_9,\\
&\ket{v_{10}}-\ket{v_{9}} =   -\frac{Q }{(\lambda^2 -1)  q \epsilon } e_9 \label{eq:v10}
\end{align}
Now,  $\ket{v_{11,12,\ldots 16}}$ are given by 
\begin{align}
&\ket{v_{11}},\ket{v_{12}} =e_{12} +  \sum_{j=2,3,8} d_j(Q) e_j,  \\
&\ket{v_{13}} =e_{2} + \frac{Q-\lambda  \left(q^2-1\right) (\epsilon -1)}{2 \left(\lambda ^2-1\right) q \epsilon } \ e_{3}\\
&\ket{v_{14}} - \ket{v_{13}}= \frac{Q}{q \epsilon(1 -\lambda ^2)  } \ e_3  \label{eq:v14}\\
&\ket{v_{15}},\ket{v_{16}} =e_{15} +  \sum_{j=5,9,14} f_j(Q) e_j \label{eq:v16}
\end{align}
where $f_j$ are some  rational expressions containing $Q$ which explicit form will not be needed in the following. 
Explicit form of the left eigenvectors 
$ \bra {w_j}$ can be written as well.  With explicit sets $\ket{v_{j}}$ and $ \bra {w_j}$ we are 
in principle able to calculate fully analytically,  a time evolution for an arbitrary 
initial state $\rho[0]$.
 
Below we shall just give information regarding the chosen observable.  

To proceed let us denote by $ \hat \bE_j $ an operator, vectorized version of which 
is $\ket{ \hat \bE_j} =e_j$.
From (\ref{eq:v10}), (\ref{eq:v14}),  we find that among simplest  observables $\bE_{j}$,  the two observables
\begin{equation}
\hat \bE_3 = \frac12 \si_1^{-} (I +\si_2^z),\quad\quad
\hat \bE_9 = \frac12  \si_1^{+} (I +\si_2^z) =  \bE_3^\dagger,
\end{equation}
are most  sensitive  to the presence of EP,  since they have an overlap with only EP-containing eigenvectors.   
Indeed,  from 
easily verifiable 
$Tr(\bE_3 \bE_j)= \de_{j,9}$,  $Tr(\bE_9 \bE_j)= \de_{j,3}$,
and inspecting (\ref{eq:v1})-(\ref{eq:v16}),
we find that 
$\bE_9$ has nontrivial overlap  with $\ket{v_{11}} ,\ket{v_{12}},\ket{v_{13}},\ket{v_{14}}$ only,  all of which 
are EP-containing eigenvectors.
Likewise,  $\bE_3$ has nontrivial overlap only with $\ket{v_{9}},\ket{v_{10}},\ket{v_{15}},\ket{v_{16}}$. 

Let us choose  an observable $\bE_3$ for definiteness. 
From (\ref{app:ObservableEevolution}),  we have 

\begin{align}
&\langle \hat e_3[n] \rangle =  \sum_{j=9,10,15,16}  (\mu_j)^n \braket{w_j}{\rho[0]} \braket{\hat e_3}{v_j}
 \label{eq:Y3Evolution}
\end{align}
(note that $\braket{\hat e_3}{e_j} = \de_{j,9}$  ).
To calculate the above, we need explicit form of $ \bra {w_j}$ for $j=9,10,15,16$.  
We find 
 \begin{align}
& \bra{w_{15}}= \frac{q (\la^2 - 1)}{Q}  \left(  e_{14} -f_{-}e_{15} \right),\\
&\bra{w_{16}}= -\frac{q (\la^2 - 1)}{Q}  \left( e_{14} -f_{+}e_{15} \right),\\
&\left(1+\frac{f_{-}^2}{\eps} \right)    \bra{w_9} = e_5 + f_{-}\,  e_9 + X_{14} e_{14}+  X_{15} e_{15},\\
&\left(1+\frac{f_{+}^2}{\eps} \right)   \bra{w_{10}} = e_5 + f_{+} \, e_9 + Y_{14} e_{14}+  Y_{15} e_{15},\\
&f_{\pm} = \frac{\lambda  \left(q^2-1\right) (\epsilon -1)\pm Q}{2 \left(\lambda
   ^2-1\right) q},
\end{align}
where $X_{14}, X_{15}, Y_{14}, Y_{15}$ are coefficients given by some complicated ratios,  explicit form of which will not be needed in the following. To simplify the calculus further,   we  choose  $\rho[0]$ to be a pure state 
of the form 
\begin{align}
&\rho[0] = \ket{\psi}\bra{\psi}, \quad \bra{\psi}  = \frac{1}{\sqrt{2}}  \bra{1,0,1,0}  \label{eq:InitialState}
\end{align}
which has no overlaps with $w_{15},w_{16}$:
 $(w_{15}, \ket{\rho[0]}) =  (w_{16}, \ket{\rho[0]})=0 $.  Then  the summation in (\ref{eq:Y3Evolution})
extends over $j=9,10$ only.
Denoting $ \braket{w_j}{\rho[0]} \braket{\hat e_3}{v_j} \equiv \ga_j$, we find 
\begin{align}
&\langle \hat e_3[n] \rangle =  \sum_{j=9,10,15,16}  (\mu_j)^n  \ga_j \\
&\ga_{15}=\ga_{16}=0,\\
&\{ \ga_9,\ga_{10} \}= \left\{  
\frac{g_{-}}{2(4 + {g_{-}}) }, \  \frac{g_{+}}{2(4 + {g_{+}}) }
 \right\}\\
&g_{\pm} = \frac{\left(\lambda  \left(q^2-1\right) (\epsilon
   -1) \pm Q\right)^2}{\left(\lambda ^2-1\right)^2 q^2 \epsilon
   }= 4\eps^{-1} (f_{\pm})^2, \label{app:gpm}
\end{align}
i.e.  Eqs.~\ref{eq:e3evolution}-\ref{eq:gpm} of the main text.

\end{document}